\documentstyle[12pt]{article}

\begin{document}

\input{epsf}

\def\beq{\begin{equation}}
\def\eeq{\end{equation}}
\def\bea{\begin{eqnarray}}
\def\eea{\end{eqnarray}}
\def\beas{\begin{eqnarray*}}
\def\eeas{\end{eqnarray*}}
\def\ov{\overline}
\def\ot{\otimes}

\newcommand{\hf}{\mbox{$\frac{1}{2}$}}
\def\sig{\sigma}
\def\De{\Delta}
\def\af{\alpha}
\def\be{\beta}
\def\la{\lambda}
\def\ga{\gamma}
\def\ep{\epsilon}
\def\vep{\varepsilon}
\def\half{\frac{1}{2}}
\def\third{\frac{1}{3}}
\def\fth{\frac{1}{4}}
\def\sth{\frac{1}{6}}
\def\tth{\frac{1}{24}}
\def\tde{\frac{3}{2}}

\def\zb{{\bar z}} 
\def\psib{{\bar \psi}} 
\def\etab{{\bar \eta }}
\def\gab{{\bar \ga}}
\def\vev#1{\langle #1 \rangle}
\def\inv#1{{1 \over #1}}

\def\CA{{\cal A}}       \def\CB{{\cal B}}       \def\CC{{\cal C}}
\def\CD{{\cal D}}       \def\CE{{\cal E}}       \def\CF{{\cal F}}
\def\CG{{\cal G}}       \def\CH{{\cal H}}       \def\CI{{\cal J}}
\def\CJ{{\cal J}}       \def\CK{{\cal K}}       \def\CL{{\cal L}}
\def\CM{{\cal M}}       \def\CN{{\cal N}}       \def\CO{{\cal O}}
\def\CP{{\cal P}}       \def\CQ{{\cal Q}}       \def\CR{{\cal R}}
\def\CS{{\cal S}}       \def\CT{{\cal T}}       \def\CU{{\cal U}}
\def\CV{{\cal V}}       \def\CW{{\cal W}}       \def\CX{{\cal X}}
\def\CY{{\cal Y}}       \def\CZ{{\cal Z}}

\newcommand{\np}{Nucl. Phys.}
\newcommand{\pl}{Phys. Lett.}
\newcommand{\prl}{Phys. Rev. Lett.}
\newcommand{\cmp}{Commun. Math. Phys.}
\newcommand{\jmp}{J. Math. Phys.}
\newcommand{\jpamg}{J. Phys. {\bf A}: Math. Gen.}
\newcommand{\lmp}{Lett. Math. Phys.}
\newcommand{\ptp}{Prog. Theor. Phys.}

\newif\ifbbB\bbBfalse                
\bbBtrue                             

\ifbbB   
 \message{If you do not have msbm (blackboard bold) fonts,}
 \message{change the option at the top of the text file.}
 \font\blackboard=msbm10 
 \font\blackboards=msbm7 \font\blackboardss=msbm5
 \newfam\black \textfont\black=\blackboard
 \scriptfont\black=\blackboards \scriptscriptfont\black=\blackboardss
 \def\Bbb#1{{\fam\black\relax#1}}
\else
 \def\Bbb{\bf}
\fi

\def\bC{{\Bbb C}} 
\def\bZ{{\Bbb Z}}
\def\CN{{\cal N}}

\title{Exact integrability of the $su(n)$ Hubbard model}
\author{{\bf Z. Maassarani}\thanks{Work supported by NSERC 
(Canada) and FCAR (Qu\'ebec).} \\
\\
{\small D\'epartement de Physique, Pav. A-Vachon}\\
{\small Universit\'e Laval,  Ste Foy, Qc,  
G1K 7P4 Canada}\thanks{email address: zmaassar@phy.ulaval.ca} \\}
\date{}
\maketitle

\begin{abstract}
The bosonic $su(n)$ Hubbard model was recently introduced. The model was shown 
to be integrable in one dimension  by exhibiting the 
infinite set of conserved quantities. 
I derive the $R$-matrix and use it to show  that the conserved charges
commute among themselves. This new  matrix is a non-additive  solution
of the Yang-Baxter equation. Some properties of this matrix are derived.  
 
\end{abstract}

\vspace*{2.5cm}
\noindent
\hspace*{1cm} PACS numbers: 75.10.-b, 75.10.Jm, 75.10.Lp\hfill\\
\hspace*{1cm} Key words: Hubbard model, $su(n)$ spin-chain, integrability

\vspace*{2.5cm}
\noindent
\hspace*{1cm} October $6^{\rm th}$ 1997\hfill\\
\hspace*{1cm} LAVAL-PHY-25/97\hfill\\

\thispagestyle{empty}

\newpage

\setcounter{page}{1}

\section{Introduction}
 
The two-dimensional Hubbard model  was introduced to
describe the effects of correlation for $d$-electrons
in transition metals \cite{guhu}. 
It was  then shown to be relevant to the study of 
high-$T_c$ superconductivity of cuprate compounds.

In one dimension  the model is integrable \cite{liwu,sh12,woa}.
The integrability framework of the model is the quantum inverse scattering 
method \cite{qism}. However, despite sharing many properties with
discrete quantum integrable models, the model has a   
peculiar integrable structure which defines  a class of its own.

In seeking to generalize the Hubbard model in any dimension, 
it was  therefore natural to  look for a one-dimensional generalization
which is integrable. 
An $n$-state  generalized  
model which contains the usual $su(2)$ model was recently introduced in
\cite{hn}. This $su(n)$ Hubbard model was shown to possess an infinite set of 
conserved charges and to have an extended $su(n)$ symmetry.  
The model is built by coupling two copies of the recently discovered
$su(n)$ XX `free-fermions' model \cite{mm}. For $n=2$ a fermionic
formulation exists, but for $n > 2$ finding an analogous framework 
is a tantalizing problem.

In this work I derive the $R$-matrix of the model; this provides a direct 
proof of the commutation of the conserved charges among themselves.
Section two gives the definition of  the bosonic Hamiltonian
and the transfer matrix.
The $R$-matrix intertwining the monodromy matrices is derived 
in section three. In section four some properties of this new matrix
are given. I conclude with some remarks and outline some outstanding issues. 

\section{The model}

Let $E^{\af\be}$ be the $n\times n$ matrix with a one at row $\af$ 
and column $\be$ and zeros otherwise. 
The $su(n)$ Hubbard Hamiltonian on a ring then reads \cite{hn}:
\bea
H_2 &=&\sum_i h_{ii+1} +\sum_i h^{'}_{ii+1} + U\sum_i h^c_i\label{h2}\\
&=& \sum_i \sum_{\af < n} \left(x E_i^{\af n} E_{i+1}^{n\af} + x^{-1}  
E_i^{n\af} E_{i+1}^{\af n} + (E\rightarrow E^{'})\right) + U
\sum_i (\rho_i +\frac{n-2}{2})  (\rho^{'}_i +\frac{n-2}{2})\nonumber
\eea
where $\rho = \sum_{\af < n} E^{\af\af} -(n-1) E^{nn}$, and 
primed and unprimed quantities correspond to two  commuting   
copies of the $E$ matrices.   
The Hamiltonians $h$ and $h^{'}$
are  $su(n)$ XX Hamiltonians \cite{mm}.
The complex  free parameter $x$ is a deformation 
inherited from the XX model. The Hamiltonian $H_2$ is defined
in one dimension but can be evidently defined on any lattice; integrability is
lost however.

For $n=2$ and $x=1$, and using Pauli matrices, the Hamiltonian is just the 
integrable bosonic version of the usual Hubbard Hamiltonian \cite{sh12}:
\beq
H_2^{(2)}=\frac{1}{2} \sum_i (\sigma^x_i \sigma^x_{i+1} + \sigma^y_i
\sigma^y_{i+1}) 
+ (\sigma\rightarrow \sigma^{'} ) +  U\sum_i \sigma^z_i\sigma^{'z}_i \nonumber
\eeq
The  Hamiltonians   can be written simply  
in terms of  $su(n)$ hermitian traceless matrices.
For $|x|=1$ the Hamiltonians are hermitian. 

The transfer matrix is the generator of the infinite set 
of conserved quantities. Its construction was given in \cite{hn}.
We recall it here.
Consider first the $R$-matrix of the $su(n)$ XX model \cite{mm}:
\bea
R(\lambda) &=&\phantom{+} a(\lambda) \; 
[E^{nn}\otimes E^{nn}+\sum_{{\af, \be<n}}
E^{\be \af}\otimes E^{\af\be}] \nonumber\\
& & + b(\lambda)\;\sum_{{\af<n}}(x E^{nn}\otimes E^{\af\af}   + x^{-1} E^{\af\af}\otimes E^{nn})\nonumber\\
& & + c(\lambda)\; \sum_{{\af<n}}(E^{n
\af}\otimes E^{\af n} + E^{\af n}\otimes E^{n\af}) 
\eea
where $a(\la)=\cos(\la)$, $b=\sin(\la)$ and $c(\la)=1$.
The functions $a$, $b$ and $c$ satisfy the `free-fermion' condition:
$a^2 +b^2 = c^2$. For this set of parameters,  a Jordan-Wigner
transformation turns the $U=0$ Hamiltonian density for $su(2)$ into
a fermionic expression for free fermions hopping on the lattice.
  
Consider also the matrix 
\beq
I_0 (h)  =\cosh (\frac{h}{2})\; {\rm Id} + \sinh (\frac{h}{2}) \; C_0 \,C^{'}_0 
=\exp\left(\frac{h}{2}\, C_0 \,C^{'}_0\right)
\eeq
where $C=\sum_{\af < n} E^{\af\af}-E^{nn}$. 
We stress  that $C$ turns out to be  the fundamental matrix, not 
the $su(n)$ generator $\rho$.
We have $\rho +\frac{n-2}{2}{\rm Id} = \frac{n}{2} C$, for $n \geq 2$.
The parameter
$h$ is related to the spectral parameter $\lambda$ by 
\beq
\sinh (2h) = \frac{n^2 U}{4} \sin (2\la) \label{rela}
\eeq
One chooses for $h(\la)$ the principal branch which vanishes for 
vanishing $\la$ or $U$.  Then for $U=0$  the monodromy matrix 
becomes a tensor product of two uncoupled XX models.
The Lax operator at site $i$ is given by:
\beq
L_{0i} (\la) = I_0(h)\, R_{0i}(\la)\, R^{'}_{0i}(\la)\, I_0(h)
\eeq
and the monodromy matrix is a product of Lax operators,
$T(\lambda)= L_{0M}(\la)...L_{01}(\la)$,
where $M$ is the number of sites on the chain. 
The transfer matrix is the trace of the monodromy matrix over 
the auxiliary space 0:
$\tau (\la)= {\rm Tr}_0 \;\left[\left( L_{0M}...L_{01}\right)(\la)\right]$.
One possible set of conserved quantities is  given by
\beq
H_{p+1} = \left({d^p \ln\tau (\la)\over d\lambda^p}\right)_{\la=0}
\eeq
The proof that  $H_2$ commutes with  $\tau(\lambda)$ was given in \cite{hn}.
The derivative of the matrix $I$ gives the coupling term appearing
in (\ref{h2}). Note that the definition involving a logarithm
has two benefits. Besides giving the most local operators, it further
disentangles the two copies. 

\section{The $R$-matrix}

We derive  the $R$-matrix intertwining two  monodromy
matrices at different spectral parameters. To this end we 
generalize the  
algebraic method of the Decorated Star Triangle Equation
introduced by Shastry \cite{sh3}.

The XX $R$-matrix satisfies the regularity
property
$\check R (0) = {\rm Id}$,  the unitarity condition
$\check R (\lambda) \check R (-\lambda) = {\rm Id} \;\cos^2\lambda$
and  the Yang-Baxter equation
\beq
\check{R}_{12}(\lambda -\mu)  R_{13}(\lambda)R_{23}(\mu) =
R_{13}(\mu) R_{23}(\lambda)\check{R}_{12}(\lambda -\mu)\label{ybec}
\eeq
where $R=P \check{R}$ and $P$ is the permutation operator
on the tensor product of   two $n$-dimensional spaces. 
It is easy to verify that it also satisfies a decorated Yang-Baxter equation
\beq
\check{R}_{12}(\lambda +\mu)\, C_1\, R_{13}(\lambda)\, R_{23}(\mu) =
R_{13}(\mu)\, R_{23}(\lambda)\,C_2\,\check{R}_{12}(\lambda +\mu)\label{dybec}
\eeq

We now look for the $R$-matrix intertwining two $L$-matrices:
\beq
\check{R}(\la_1,\la_2) \; L(\la_1)\otimes L(\la_2) =  L(\la_2)\otimes L(\la_1)
\;\check{R}(\la_1,\la_2)\label{rll}
\eeq 
The $su(2)$ case lead us to consider the following Ansatz \cite{sh3}:
\bea
\check{R}(\la_1,\la_2) &=& I_{12}(h_2) I_{34}(h_1) \left(
\alpha \,\check{R}_{13}(\la_1-\la_2)  \check{R}_{24}(\la_1-\la_2) 
\right.\nonumber\\
& &\left.+\beta\,  \check{R}_{13}(\la_1+\la_2) C_1 
\check{R}_{24}(\la_1+\la_2)C_2 \right) I_{12}(-h_1) I_{34}(-h_2)
\eea
The $R$-matrix acts on the product of four auxiliary spaces labeled from
1 to 4, and $\alpha$, $\beta$ are to be determined.
One then requires relation (\ref{rll}) to be satisfied and 
uses (\ref{ybec}) and (\ref{dybec})  to derive the following equation:
\beas
&\left(\alpha \,\check{R}_{13}(\la_1-\la_2)  \check{R}_{24}(\la_1-\la_2) 
+\beta\,  C_3\check{R}_{13}(\la_1+\la_2)C_4 \check{R}_{24}(\la_1+\la_2)
\right)I_{12}(2h_1) I_{34}(2h_2) =&\\
&I_{12}(2h_2) I_{34}(2h_1) \left(
\alpha \,\check{R}_{13}(\la_1-\la_2)  \check{R}_{24}(\la_1-\la_2) 
+\beta\,  \check{R}_{13}(\la_1+\la_2) C_1 \check{R}_{24}(\la_1+\la_2)C_2
\right)& 
\eeas
Expanding the exponentials and taking into account all the terms 
yield only two equations:
\beq
\frac{\beta}{\alpha}= \frac{b}{B}\tanh(h_1+h_2)
\;\;, \;\;\; \;\;
\frac{\beta}{\alpha}= \frac{a}{A}\tanh(h_1-h_2)
\eeq
where $a=\cos(\la_1-\la_2)$, $b=\sin(\la_1-\la_2)$, $A=\cos(\la_1+\la_2)$
and $B=\sin(\la_1+\la_2)$.
The compatibility equation
\beq
\frac{\tan(\la_1-\la_2)}{\tan(\la_1+\la_2)} = \frac{\tanh(h_1-h_2)}{\tanh(h_1+h_2)} 
\eeq 
is satisfied provided
equation (\ref{rela}) is satisfied for the pairs $(\la_1,h_1)$ and
$(\la_2,h_2)$. 
One can then  pull out $\alpha=\alpha(\la_1,\la_2)$ 
which appears as an arbitrary normalization of the $R$-matrix, to obtain:
\bea
\check{R}(\la_1,\la_2)&=& \alpha(\la_1,\la_2) I_{12}(h_2) I_{34}(h_1) \left(
\check{R}_{13}(\la_1-\la_2)  \check{R}_{24}(\la_1-\la_2)
+\frac{\sin(\la_1-\la_2)}{\sin(\la_1+\la_2)}\right.\nonumber\\
& \times&\left. \tanh(h_1+h_2) 
\check{R}_{13}(\la_1+\la_2) C_1 \check{R}_{24}(\la_1+\la_2)C_2 
\right) I_{12}(-h_1) I_{34}(-h_2)\label{rc}
\eea

The monodromy matrix being a tensor product 
of $M$ copies of $L$ matrices, one has 
\beq
\check{R}(\la_1,\la_2) \; T(\la_1)\otimes T(\la_2) =  T(\la_2)\otimes T(\la_1)
\;\check{R}(\la_1,\la_2)\label{rtt}
\eeq
Taking the trace over the auxiliary spaces and using 
the cyclicity property of the trace one obtains $[\tau(\la_1),\tau(\la_2) ]=0$.
We have thus proven that all the conserved charges $H_p$ mutually commute.

Note that this proof is rigorous and valid {\it for all values of} $n$,
and for arbitrary values of the complex parameter $x$. 
It only  involves  the algebraic properties of the operators appearing
in the various matrices, not the specific $n$-dependent representation.
The equations  (\ref{ybec}) and (\ref{dybec}) are the only equations
of this type needed for the proof.

\section{Properties of the $\check{R}$ matrix}

I now  give  some properties of the $R$-matrix. At $U=0$ the two 
$XX$ models decouple and $h(\la,U)=0$. Expression (\ref{rc})
indeed  decouples as a tensor product of two $su(n)$ XX 
$\check{R}$-matrices.

The matrix also satisfies the regularity property 
\beq
\check{R}(\la_1,\la_1) = \alpha(\la_1,\la_1)\; {\rm Id}
\eeq
and the unitarity property:
\bea
\check{R}(\la_1,\la_2) \check{R}(\la_2,\la_1)& =& \alpha^2(\la_1,\la_2)
\cos^2(\la_1-\la_2)\nonumber\\
&\times&\left(\cos^2(\la_1-\la_2) -
\cos^2(\la_1+\la_2) \tanh^2(h_1-h_2)\right) {\rm Id}
\eea
The derivation of the  last property is straightforward and involves 
algebraic relation between the building 
blocks of the $su(n)$ XX $\check{R}$-matrix.

One can invoke the associativity of the algebra of $L$ matrices, which 
ultimately reduces to the associativity of usual matrix multiplication,
to conclude that the intertwiner satisfies a Yang-Baxter relation 
of its own.  The two ways of permuting a product of three $L$-matrices
imply
\bea
&\check{\Pi} \;L(\la_1)\otimes L(\la_2)\otimes L(\la_3)\; \check{\Pi}^{-1}=
L(\la_1)\otimes L(\la_2) \otimes L(\la_3) &\nonumber\\
&\check{\Pi} = \left( \check{R}_{12}(\la_2,\la_3) \check{R}_{23}(\la_1,\la_3)
\check{R}_{12}(\la_1,\la_2) \right)^{-1} \check{R}_{23}(\la_1,\la_2) 
\check{R}_{12}(\la_1,\la_3) \check{R}_{23}(\la_2,\la_3)
\eea
I am unaware of the existence  of an equivalent of the 
Schur lemma for the algebra of $L$-matrices. This would allow to conclude
that $\check{\Pi}\propto {\rm Id}$. Once proportionality is 
established, the regularity property ensures that the proportionality constant
is one.  
We can argue 
that  $\check{\Pi} =  {\rm Id}$ holds because it has been explicitly 
verified for $n=2$  \cite{sw},  and because the building blocks
of the matrix satisfy  algebraic relations which are 
independent of $n$.
Thus the $R$-matrix satisfies the Yang-Baxter equation:
\beq
\check{R}_{12}(\la_2,\la_3) \check{R}_{23}(\la_1,\la_3)
\check{R}_{12}(\la_1,\la_2) = \check{R}_{23}(\la_1,\la_2) 
\check{R}_{12}(\la_1,\la_3) \check{R}_{23}(\la_2,\la_3)\label{rybe}
\eeq
where $\la$ and $h$ are related through (\ref{rela}).

Using the explicit expression (\ref{rc}), it should be possible
and it is instructive  to try to check that 
the above equation is satisfied for any value of $n$. 
The factors $I$ drop out  and half of the terms on both sides 
of the YBE  compensate each other because
relations (\ref{ybec}) and (\ref{dybec}) hold. The eight remaining terms 
involve highly non-trivial relations. Each term is a product of six 
$R$-matrices, three for every copy, in the ordering dictated by the YBE.
The $C$ factors can be dropped by changing the arguments of the $R$-matrices
appropriately. However the arguments do not allow the use of (\ref{ybec},
\ref{dybec})
because the middle argument is not the sum of the extreme ones. One 
is forced to expand all the products on a basis and to regroup
terms and check that the resulting trigonometric constraints are satisfied.  
I have not verified
whether all these relations hold. Although specific particularities
pertaining to the XX matrix are needed, I stress again that the proof 
is algebraic. In this respect the proof of \cite{sw} for $n=2$, 
although following a different approach, should generalize in 
a straightforward way to any value of $n$.

\section{Conclusion}

We have shown that all the conserved charges of the $su(n)$ 
Hubbard model mutually commute by exhibiting
the intertwining matrix.  This matrix is  the $su(n)$ generalization
of the $su(2)$ one obtained in \cite{sh12}. Some properties where then derived.
A notable feature of the matrix is its non-additivity property; the 
$\la$ dependence cannot be reduced to a difference $(\la_1-\la_2)$.

One can now  start diagonalizing the Hamiltonians
by the method of the algebraic Bethe Ansatz. Preliminary results
suggest an interesting structure for the Bethe eigenstates \cite{win}.

One can also consider the extension of the $su(n)$ model to other
algebras. The algebraic underlying structure of the  $su(n)$ XX model
should admit generalizations \cite{mm,win}.

\bigskip\ {\bf Acknowledgement:} I thank P. Mathieu for
bringing to my attention reference \cite{sh3} and for continued support.

\bigskip\ {\bf Note:} While this work was being written, 
Martins exhibited a gauge-transformed version of the foregoing 
$\check{R}(\la_1,\la_2)$ matrix. The derivation is also based
on a generalization of Shastry's method. However, the proof 
in \cite{mjm} was carried out for $n=3$, and `extensive checks' were
made for $n=3,4$.  The expression of $\check{R}$  for all values
of $n$ is left as a (correct) {\it conjecture}. 
This reference also used unnecessarily complicated
versions of the  Yang-Baxter equations.

\end{document}